\documentclass%
[prl,amsmath,showpacs,twocolumn,preprintnumbers,superscriptaddress]%
{revtex4}

\usepackage{mathptmx}

\usepackage{microtype}

\usepackage{graphicx}
\usepackage{dcolumn}
\usepackage{amsmath}
\usepackage{mathtools}
\usepackage{bm}

\usepackage{mathptmx}
\usepackage{microtype}

\usepackage{graphicx}
\usepackage{dcolumn}
\usepackage{amsmath}
\usepackage{mathtools}
\usepackage{bm}
\usepackage{amssymb}

\begin{document}

\title{Can the periodic spectral modulations of the 236 SETI candidate from Sloan Sky Survey stars be due to Dark Matter effects?}

\author{Fabrizio Tamburini} 
\address{ZKM -- Zentrum f\"ur Kunst und Medientechnologie, Lorentzstr. 19, D-76135, Karlsruhe, Germany.}
\address{MSC -- bw,  Stuttgart, Nobelstr. 19, 70569 Stuttgart, Germany.}

\author{Ignazio Licata}
\address{Institute for Scientific Methodology (ISEM) Palermo Italy.}
\address{School of Advanced International Studies on Theoretical and Nonlinear Methodologies of Physics, Bari, I-70124, Italy.}
\address{International Institute for Applicable Mathematics and Information Sciences (IIAMIS), B.M. Birla Science Centre, Adarsh Nagar, Hyderabad -- 500 463, India.}

\begin{abstract} 
The search for dark matter (DM) is one of the most active and challenging areas of current research.
Possible DM candidates are ultralight fields such as axions and weak interacting massive particles (WIMPs).
Axions piled up in the center of stars are supposed to generate matter/DM configurations with oscillating geometries at a very rapid frequency, which is a multiple of the axion mass $m_B$ [Brito {\it et~al.} (2015); Brito {\it et~al.} (2016)].
Borra and Trottier (2016) recently found peculiar ultrafast periodic spectral modulations in $236$ main sequence stars in the sample of $2.5$ million spectra of galactic halo stars of the Sloan Digital Sky Survey ($\sim 1 \%$ of main sequence stars in the F--K spectral range) that were interpreted as optical signals from extraterrestrial civilizations, 
%%% cambiato
suggesting them as 
possible candidates for the search for extraterrestrial intelligence (SETI) program.
We argue, instead, that this could be the first indirect evidence of bosonic axion-like DM fields inside main sequence stars, with a stable radiative nucleus, where a stable DM core can be hosted.
These oscillations were not observed in earlier stellar spectral classes probably because of the impossibility of starting a stable oscillatory regime due to the presence of chaotic motions in their convective nuclei.
The axion mass values, $(50 < m_B < 2.4 \times 10^{3})~ \mathrm{\mu eV}$, obtained from the frequency range observed by Borra and Trottier, $(0.6070< f <0.6077$) THz, agree with the recent theoretical results from high-temperature lattice quantum chromodynamics 
[Borsanyi {\it et~al.} (2016); Borsanyi {\it et~al.} (2016b)].
%\cite{qcd,qcd2}.
\end {abstract}
%\pacs{95.35.+d, 14.80.Va, 97.10.Cv, 97.10.Sj}
\maketitle
%%%%%%%%%%%%%%%%%%

%\begin{keywords}
%dark matter -- oscillations -- extraterrestrial intelligence
%\end{keywords}

\section{Introduction}

Since Zwicky (1933) pointed out the discrepancy between the dynamical and luminous matter in the Coma cluster of galaxies stressing the need for some form of Dark Matter, the progress in our knowledge both in data collecting, analysis and theory (especially in the last 40 years) has been tremendous. Dark Matter (DM) in the realm of General Relativity and conventional cosmology is nowadays a fact even if the community remains open to fundamental changes in the theory (MOND  or emergent theories of gravity for instance)  \cite{mond,exiri,verlinde11,verlinde16,licata}. 
The most compelling evidence for the existence of DM is that the gravitational and luminous matter signatures are clearly separated in many cases, like in galaxy rotation curves and gravitational lensing. DM is supposed to play a main role in galaxy formation through the gravitational effects of DM halos and in galaxy evolution, in the formation of the anisotropies of the cosmic microwave background and also explain the formation of the large scale structures observed in the Universe.
Perhaps one of the best evidence of its reality is the Bullet cluster \cite{clowe,mark} and the clusters alike. Here we are witnessing the collision of two clusters of galaxies where the collisionless DM, as evidenced by the weak lensing observations, remains unperturbed, while the gas (detected in the Xray), is highly shocked.

Concordance Cosmology following the latest optical and cosmic microwave background observations call for the following parameters: DM is responsible for the $26.8\%$ of the total mass-energy of the Universe, whilst ordinary matter is only $4.9\%$ \cite{planck}. Clearly gravitation in the Universe is dominated by DM so that the formation and evolution of cosmic objects and large structures is guided by DM.
It is therefore of paramount importance, for our knowledge of physics and cosmology to understand what we are dealing with and to collect all those hints that may lead us to a good understanding. This will lead us to know the characteristics of the particles and eventually lead to detection.

From Big Bang nucleosynthesis \cite{frenk,field}, DM is thought to be made up of exotic invisible baryonic and non-baryonic matter, such as axion-like light fields  \cite{duffy,peccei,wil,sikvie}, neutralinos and other particles from SUSY theories. Other candidates are sterile or massive neutrinos and weakly interacting massive particles (WIMPs)  \cite{bertone}. 
More stringent limits to WIMP's mass and cross section in spin-independent elastic WIMP-nucleon interactions were recently set after their 
non-detection in LUX and PandaX-II \cite{lux,panda} experiments. 
For a review on DM, see Ref. \cite{dm}.

If DM were made mostly of ultralight fields, they are expected to be axions or axion-like candidates,
cold and weakly interacting particles - either scalars or vectors - with a very small rest mass \cite{dm2}. 
Axions are pseudo-Goldstone bosons that were introduced in the Peccei--Quinn \cite{peccei,kolbe} symmetry to resolve the strong charge--parity (CP) problem of quantum chromodynamics. In this theory the expected mass of the axion $m_B$ is a free parameter, $10^{-6}~\mathrm{\mu eV}< m_B<1~\mathrm{MeV}$, currently under experimental validation
\cite{battesti,beck,cooray,admx,graham,rosenberg}.
Recently, Borsanyi \textit{et~al.} obtained a more stringent mass range $50 - 1500 ~\mu eV/c^2$ from numerical simulations of lattice quantum chromodynamics (QCD) by calculating the equation of state of QCD and the topological susceptibility for low up to very high temperatures including also Standard Model particles in an Euclidean space-time lattice \cite{qcd,qcd2}. 

\section{Dark Matter effects on stars.}
The possible influence of DM in the evolution and structure of stars has been widely discussed in the literature. 
Depending on the type of DM particles considered, one can find slightly different effects that may alter the structure and evolution of a star such as the increase of energy production or, conversely, the dissipation of the internal energy at a faster rate. 
To give an example, the presence of an high concentration of WIMPs accumulated in the core of a star is supposed to trigger self-annihilation processes between WIMPs, becoming an additional source of energy \cite{scott}, causing the decreasing of temperature and pressure in the stellar core and increasing the life of all the stellar spectral classes,  with a more significant contribution to lower mass stars \cite{sulisti}.
Light DM particles, instead, can shorten the lifetime to all stars transporting away the energy generated by the nuclear reactions in a very efficient way \cite{kolbe}. Because of their small cross sections, axions and ultralight Axion-like DM particles may give rise to non-standard cooling mechanisms, similar to the neutrino cooling mechanism. This axion cooling mechanism is used to explain some discrepancies found between the observational data and the standard theoretical models of stellar evolution \cite{kolbe,giannotti}.

Bosonic DM may also perturb the stability of the structure of a star. The most interesting effect induced by ultralight bosonic DM fields is the onset of ultrafast oscillations of stellar configurations described by a perfect fluid and a bosonic condensate.
In fact, real scalar massive bosonic fields minimally coupled to gravity can have a non-trivial time dependent stress-energy tensor that gives rise to long-term stable oscillating geometries: self-gravitating bosonic fields such as axions or axion-like fields can form particular ``breathing'' configurations, where both the spacetime geometry and the field oscillate at a frequency that depends on the mass of the boson.

Stable DM cores can form inside stars. DM can accrete and cluster inside compact and non-compact stars and form oscillating DM cores when both the bosonic DM field and the gravitationally--coupled fluid density vary periodically.
In these oscillating configurations both the stellar material and the field oscillate at a precise multiple of a fundamental frequency $f_m$ that depends on the mass $m_B$ of the boson,
\begin{equation}
f_m=2.5 \times 10^{14} ~ \frac{m_B c^2}{eV} ~ Hz.
\label{eq1}
\end{equation}
These results, obtained with a simple model that describes the star as a regular fluid sphere with an equation of state of polytropic index $n$ and the usual pressure-density relationship $P \propto \rho^{(n+1)/n}$, remain valid also for models described by a more general equation of state, where the oscillating components in the equation of motion behave as small perturbations of a static star \cite{brito,brito2}.

\section{Star spectral modulation and axion mass.}

Here we make the \textit{ansatz} of extending these results also to the inner core of lower spectral classes main sequence (MS) stars where the energy flux in the inner core is carried out by radiative transfer. This region of the star, also known as radiative zone, can be described in terms of a static fluid.
Following Brito \textit{et al.} \cite{brito,brito2}, we argue that the local density of the radiative stellar core, when enough DM is piled up in the center, becomes a periodic function of time with a period given by the properties of the scalar field.
The two fluids of ordinary and dark matter oscillate together with spacetime, with the effect of transmitting this ultrafast oscillatory motion outwards to the outer layers of the star, with experimentally observable effects.
The decrease of DM density and of energy due to the flux of axions leaving the star should generate a variation in the amplitude of the oscillations that could be likely observable.

From the theory of stellar structure and evolution, these oscillatory motions can occur in MS stars with masses lower than $1.2~M_\odot$ down to $0.3~M_\odot$  that correspond to a range of spectral classes from F to M. 
Stars with masses $M \simeq 1.2-1.3 ~M_\odot$ are almost totally radiative and, for smaller masses, the stellar core is surrounded  by a convective envelope that becomes larger and larger the smaller is the mass of the star. Main sequence stars with less than $0.3~M_\odot$ are instead totally convective. Conversely, for masses larger than $1.2~M_\odot$ the core region starts becoming convective and is surrounded by a radiative envelope, an environment that does not favor the onset of DM-induced ultrafast oscillatory motions.
In fact, the onset of a stable oscillatory regime, in these conditions, can occur only if its frequency is much lower than that of the chaotic rapid variations induced by the convective motions in the stellar core, namely, when the chaotic flows become uniformly indistinguishable from white noise and can modeled in terms of a stochastic flow coupled to a slowly-evolving deterministic flow that should describe our ultrafast oscillations \cite{denker,taylor,bt98}, and it is clearly not our case. 

% cambiamento
DM-induced oscillations could be actually the possible explanation of the spectral modulations, with periods varying from $1.6453  \times 10^{-12}$ up to $1.6474  \times 10^{-12}$ seconds, found by Borra \& Trottier  \cite{borra} in $236$ MS halo stars obtained from the Fourier transform analysis of $2.5$ million spectra from the Sloan Digital Sky Survey (SDSS). This sample corresponds to the $\sim 1 \%$ of the sub-set of metal-poor population II MS stars in the halo \cite{jofre} with a radiative nucleus, in the spectral interval mainly from F to K, for which SDSS instrumentation could reveal this type of signal. 

The observational data were taken in the spectral regions of blue (380 nm -- 600 nm) and red (600 nm -- 9200 nm)  and the SDSS spectra were obtained from two different spectrographs, one in the blue region $(380 nm <  \lambda  < 615 nm)$ and one in the red region $(580 nm <  \lambda <  920 nm)$, while the signals extend over the entire spectral range. This excludes the presence of a systematic instrumental error.

These spectral modulations could be caused by the ultrafast oscillations induced by the pulsating matter/DM cores hosted in these stars. 
In fact, objects that emit light and oscillate at those frequencies generate periodic spectral modulations detectable in their astronomical spectra, similarly to a set of high-intensity light pulses separated by constant time intervals shorter than $10^{-10}$ seconds \cite{borra2010}.

\begin{figure}[hbt]
\center
\thicklines
\includegraphics[scale=0.30]{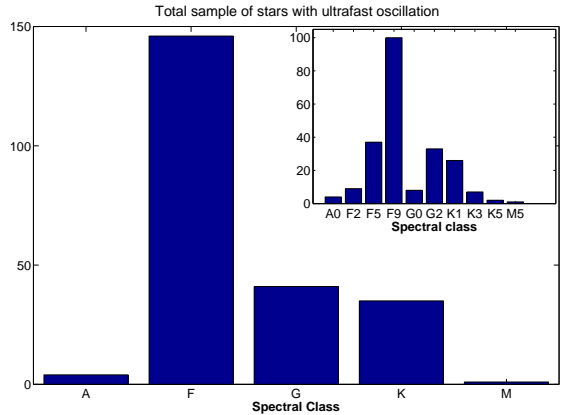}
\caption{Histogram of the spectral class of stars presenting ultrafast oscillations. The maximum peak is found for the F spectral-type class, with a peak on the F9 subclass where the energy transfer is mostly radiative and the star can be modeled as a fluid sphere. This is the optimal situation for the onset of stable dark matter ultrafast oscillations.
In general, these stars have an inner core that can behave like a fluid ball because convection is not present there and the upper layers have thinner outer convective regions that may damp the signal amplitude with respect to the lower-mass classes. The presence in the sample of four A0 and one M5 stars are briefly discussed in the text.}
\label{fig1}
\end{figure}

As shown in Fig \ref{fig1}, we remark that all these stars belong to a narrow spectral range centered near the spectral type of the Sun, a number of $234$ stars that have spectral classes mostly in the F--K spectral range with some exceptions like very small samples of A and M class stars. The distribution is peaked on the F class with a smooth decay towards the lower spectral regions where the outer convective region becomes larger and larger and would not contribute actively to the ultrafast oscillations of the radiative nucleus.
The inset of Fig \ref{fig1} shows that F9 is the dominating spectral class of this sample, while G0, F2 and K3 stars represent about the $1 \%$ of the signal and there are very few samples of A0, K3, K5 and M5 stars.
A possible explanation to the fact that the majority of the signals have been found in spectral types ranging from F to G might be also due 
to a selection effect: the SEGUE survey targeted stars are mostly F to K-type stars and, in addition to that, F-G stars are the dominant populations in our galaxy.
Noteworthy, we point out that the dominant stellar spectral type in which the distribution is peaked is the F9 spectral class that has a mass of $M \sim 1.2~M_\odot$ for which the energy transfer is almost totally radiative, in agreement with our idea: at a first approximation, in the equations of motion of the fluid, the whole star could be described as a fluid sphere, the optimal conditions for the onset of star/DM oscillations.

Stellar rotation, population type, metallicity, dark matter effects and stellar evolution processes may influence the energy transport in the inner core, with the result of modifying or even extending the mass range to higher mass values for which a star can host a stable inner core that could explain the presence, in that sample, of a four A-spectral type stars where convection in the inner core is expected and one M5 star \cite{jofre,clayton,padma,iben}. 

Our interpretation could be an alternative to the fascinating possibility that the spectral modulations are caused by high-intensity and rapidly variating pulses of light generated by extraterrestrial civilizations \cite{borra2012}, sharing a sort of common protocol based on almost identical sets of ultrafast light pulses: this ultrafast oscillatory behavior can be due to effects of axion-like dark matter.

Applying the experimental results presented in Ref. \cite{borra} to Eq. \ref{eq1}, one can easily obtain the mass $m_B$ of the axion that depends on a precise multiple of the main frequency $f_m$ dictated by the axionic field. 
For the frequency range $(0.6070 < f < 0.6077)$ THz derived from the spectral modulations, we find this mass range for the axion $(50 < m_B < 2.4 \times 10^{3})~ \mathrm{\mu eV}$.

%% cambiamento
Interestingly, 
as reported in Table 1,
this mass value overlaps the mass range found with Lattice QCD simulations, $50 - 1500 ~\mathrm{\mu eV}/c^2$ \cite{qcd,qcd2} when $2 \leq k \leq 48$ and this range is also compatible with the possible axion detections in Josephson junctions \cite{beck}, $m_B \simeq 110 ~ \mathrm{\mu eV}$ and with the results from solar observations \cite{kims}, when the frequency observed experimentally, $f$, is one of the multiples of the fundamental oscillating frequency of the boson, $f_m$, i.e. $f=k f_m$, as discussed in Refs. \cite{brito,brito2}.
In fact, for $k=1$ one obtains a mass of $m_B = 2428~\mathrm{\mu eV}$ for the lower bound of the detected frequency, $f=6.070 \times 10^{11}$ Hz, and  $m_B = 2430.8~\mathrm{\mu eV}$ for $=0.6077\times 10^{11}$ Hz. 
%%  cambiato
Larger values of $k$ correspond to lower frequencies $f_m$ and thus to smaller values of the mass of the axion. For more details, see Tab. \ref{tab1}. 
%%% cambiamento
For $k=48$, the lower mass value for the axion obtained from lattice QCD simulations \cite{qcd} is found.
This range of masses agrees also with the limits fixed by stellar evolution and the neutrino cooling of SN1987a that precludes axions in the mass range of $10^{-3}$ up to $2$ eV. These axions are non-thermal, with lifetime much
longer than the age of the Universe. In this scenario axions can be hardly detected from their decay in two photons
\cite{kolbe}.
Interestingly, the mass range we found overlaps also the mass range of the axionic DM of the Standard Model Axion Seesaw Higgs (SMASH)  \cite{balles}.

\begin{table}
\begin{center}
\begin{tabular}{|c|c|c|}
\hline
Multiplication & fundamental & axion mass \\ 
factor $k$ &  frequency $f_m$ (THz) & $\mu$eV  \\
\hline
 $1$ &   $ 0.607 $ & $ 2.4308  \times 10^{3}$  \\ \hline
 $2$ &   $ 0.304 $ & $ 1.2154 \times 10^{3}$  \\ \hline
 $3$ &   $ 0.202$ & $ 810.2667$  \\ \hline
 $4$ &   $ 0.152$ & $ 607.7000$  \\ \hline
 $8$ &   $ 0.076$ & $ 303.8500$  \\ \hline
 $16$ & $ 0.038$ & $ 151.9250$  \\ \hline
 $32$ & $ 0.019$ & $75.9625$  \\ \hline
 $48$ & $ 0.013$ & $50.6417 $  \\ \hline
\end{tabular}
\end{center}
\caption{ Axion mass, $m_B$, derived from the multiples of the main frequency oscillations $f_m$ of the axionic field. The experimentally observed frequency $f$ is considered as a multiple harmonic of a fundamental oscillation $f_m$ due to dark matter on the star, namely, $f = k f_m$ \cite{brito,brito2}. We report the range of values of $k > 1$ for which $m_B$ overlaps with the results of Lattice QCD simulations \cite{qcd}.}
\label{tab1}
\end{table}

Following Ref. \cite{borra}, there might be other different phenomena able to explain the presence of these spectral modulations such as rapid pulsations in small regions of the atmospheres of the stars. However, the period of the pulsation, which is on the order of $1.65 \times 10^{-12}$ seconds, is unrealistically small for a standard stellar structure. These rapid pulsations cannot even be ascribed to the effect of highly peculiar chemical compositions that could be present in a small fraction of galactic halo stars, as these peculiarities were not observed.
Other causes e.g. due to the difference in the strength of spectral lines, chemical composition, effective temperature, gravity, rotational velocities, turbulence or radial velocities are excluded, favoring our DM conjecture.

\section{Discussion and conclusions}

The Fourier analysis of the spectra of $2.5$ million of stars in the Sloan Sky Survey revealed peculiar spectral modulations in $236$ main sequence stars of our galactic halo  ($0.00936 \%$ of the whole sample) with spectral classes in a narrow range close to that of our Sun.
More precisely, Borra and Trottier found that the actual fraction of stars presenting these ultrafast oscillations correspond to the $\sim 1 \%$ of the sub-set of metal-poor population II main sequence stars in the halo \cite{jofre} hosting a radiative nucleus (F--K spectral types) for which SDSS could reveal this type of signal. 
After having excluded any possible origin from instrumental errors, data analysis and other standard astrophysical effects, these modulations were supposed to be due to optical signals from extraterrestrial civilizations \cite{borra}.

We argue, instead, that these spectral modulations are of astrophysical origin and are due to ultrafast oscillations of the stellar structure of these stars induced by the presence of ultralight fields such as axions or axion-like dark matter in their cores that may induce these ultrafast oscillations with frequencies proportional to the axion mass $m_B$ \cite{brito,brito2}.
Assuming that the frequency $f$ derived from the observed spectral modulations is a multiple of the fundamental oscillating frequency $f_m$ induced by the boson, we found a range of axion masses $(50 < m_B < 2.4 \times 10^{3})~\mathrm{\mu eV}$ that surprisingly overlaps the recent simulations of lattice QCD \cite{qcd,qcd2} 
when the parameter $k$ is found in the range $2 \leq k \leq 48$.

Interestingly, these oscillations occur only in certain spectral types of stars, where a hydrodynamically-stable core is present because the energy transfer occurs through radiative processes. Simple considerations suggest that this could be the most favorable environment where ultralight axionic DM hosted inside a stellar nucleus can impose a rapid oscillatory regime to the inner core and to spacetime that propagates to the outer layers of the star.
In stars with masses around $1.2~M_\odot$  (F-G stars) the energy transfer is totally radiative and the whole stellar structure can be modeled as a  fluid sphere. 
This could explain why the sample of $236$ stars is peaked around the F9 spectral class, one of the dominating spectral classes in our galaxy.
Only five stars, four A-spectral class stars, that are supposed to start to have a convective core and one M5 star that is supposed to be totally convective present the same spectral modulation. These exceptions might be caused by other astrophysical effects or even by the presence of DM fields that may alter the structure of these stars.

These stars have been observed to oscillate almost at the same frequency with similar oscillation amplitudes that would favor the the DM hypothesis, as the amplitude depends on the DM characteristics. 
The energy of these oscillations is about $10^{-5}$ the energy emitted in the SDSS spectral range \cite{borra}. For a Sun-like star, this corresponds to a power of $10^{21}$  W. 

Following Refs. \cite{brito,brito2}, the results from observations may confirm the DM scenario where is expected a dependence of the amplitude on the mass coupling of the scalar field and on the energy density of the scalar field and fluid. 
The formation of these composite stars occurs though two different channels: through gravitational collapse in a DM-rich environment, forming a star with a DM core inside, or, in the second case, from capture and accretion of DM through the so-called gravitational cooling mechanism that prevents stars growing to the collapse through the ejection of mass, even when DM is composed of light massive fields. 

To give an order of magnitude estimate of the mass coupling parameter $\mu_s $ that characterizes the oscillatory motion, let us consider an averaged value of the axion mass in the range overlapping lattice QCD simulations, $m_B \simeq 7 \times 10^{2}~\mathrm{\mu eV}$, and in the mass range of the radiative nuclei of F to K spectral type stars, one obtains $0.2 < \mu_s M < 7$ that allows stable oscillatory regimes.
%%%
For the higher values of the mass coupling parameter, localized and stable configurations of self-gravitating bosonic DM cores inside stars are possible.
%%%%
In fact, when we deal with scenarios involving mixed scalar oscillaton and fermion fluids, where the total baryon number is conserved, we find that the numerical simulations by Brito \textit{et al.} indicate that the time averaged energy density of the scalar field, $\rho_\phi$, can be very peaked at the center of the star with the same order of magnitude of the fermion fluid energy density, $\rho_F \sim 10^{-4}$. Moving far away from the center of the star, $\rho_\phi$ rapidly falls down, assuming values that can be orders of magnitude smaller than $\rho_F$.
On the other hand, smaller values of $m_B$, compatible with lattice QCD simulations, give a completely different scenario where stars present an extended scalar condensate that protrudes away from their structure, with a negligible influence on the fluid distribution. In this case stars are expected to present broad and light oscillatory regimes where the energy density of the scalar field $\rho_\phi$ at the center of the star can differ by two orders of magnitude with respect to $\rho_F$.

To explain why only $1$ over about one hundred of MS halo stars with a radiative nucleus oscillate, energy considerations derived from the experimental results by Borra and Trottier seem to favor at a first glance scenarios with small values of $m_B$ unless invoking a damping effect due to the viscosity of the stellar matter.
In fact, the oscillations are supposed to be damped by the viscosity of the stellar material that will lead to a depletion of the scalar field core and the damping 
in the outer layers of the star; in any case, the exact effect of viscosity and the local thermalization of the scalar field with the stellar material together with the effects of the stellar magnetic field in the outer regions of the stellar structure that are thought to modify the viscosity of the outer layers of the star is still not well determined.

Moreover, according to the first and second DM stellar formation channels, a non-uniform distribution of DM-rich regions in the galactic halo may act as further selection effect, when considering scenarios where axions may cluster in dense small-scale substructures such as axion miniclusters, that can tidally interact with stars, captured and disrupted forming tidal streams with densities that are one order of magnitude larger 
than the average value \cite{tynakov,hardy}.

One can also argue that the small sample of stars observed could be due to an oscillatory motion with a transient behavior due to axion loss because of interaction with matter and fields or because axions may actually leave the gravitational potential well of star. For example, axions may be influenced by gravity in a process of condensing cold axion particles and/or evaporating the axionic field; instead, axion DM appears to be stable with respect to gravity behaving as a classical field, remaining coherent and gravitationally confined by the star \cite{davidson}. In fact, under certain conditions, axions in the range of masses $\mu_B$ found here can cluster even down to clumpy configurations such as axionic star-like structures \cite{barranco} suggesting that the evaporation of the axionic field from an isolated star can be negligible as these particle mainly interact gravitationally.
Other additional effects that can decrease the number of axions in the core is the interaction of axions with matter and photons \cite{raffelt} or with the stellar magnetic fields that permeate down to the radiative zone producing photons \cite{gnedin} that are almost negligible with respect to the effects of viscosity damping but that can be revealed as an extra production of energy from the star. In MS stars, this hypothetical axion loss can be easily counterbalanced by DM accretion onto the radiative zone. In any case, a much deeper theoretical investigation is needed to understand this point.

At our knowledge, the search of such ultrafast periodic modulations in the stars of our galactic plane has not been made systematically as in the work by Borra and Trottier \cite{borra}. The study of the modulations in the spectrum of the Sun, even though not trivial because it is an extended object, may help the current investigation of the presence of DM in our star \cite{kims}.
What we need actually to support or discard our DM conjecture is an extensive analysis of the spectra of other stars in globular clusters \cite{viaux}, in the galactic plane and where high concentrations of DM are expected taking into account also the effects of different metal composition of stars \cite{redondo} where the presence of a forest of metallic spectral lines would make the detection of the oscillatory motion experimentally difficult.

\subsection*{Acknowledgments}
We acknowledge Guido Chincarini and Antonio Bianchini for helpful comments and discussions.
F.T. gratefully acknowledges the financial support from ZKM and MSC-bw.
\\

%%%%%%%%%%%%%%%%

\end{document}